\documentstyle [12pt]{article}

\parskip=\medskipamount                 
\def\7#1#2{\mathop{\null#2}\limits^{#1}}        
\def\ast{\displaystyle *}

\def\beee{\begin{equation}}
\def\eeee{\end{equation}}
\def\dggg{^{\dagger}}
\begin{document}

\bibliographystyle{unsrt}

\begin{center}
{\Large \bf VIRTUES OF THE HAAG EXPANSION IN QUANTUM FIELD THEORY}
\footnote{Supported in part by the National Science Foundation.\\
e-mail address, greenberg@umdep1.umd.edu}\\
[5mm]
O.W. Greenberg\\
{\it Center for Theoretical Physics\\
Department of Physics \\
University of Maryland\\
College Park, MD~~20742-4111}\\[5mm]

Talk given at the First Arctic
Workshop on Future Physics and Accelerators, August, 1994, Saariselka,
Finland\\
To appear in the Proceedings\\

Preprint number 95-99\\
\end{center}

\vspace{2mm}

\begin{center}
{\bf Abstract}
\end{center}

I survey the use of the Haag expansion as a technique to solve quantum field
theories.  After an exposition of the asymptotic condition and the Haag
expansion, I report the results of applying the Haag expansion to several
quantum field theories, including galilean-invariant theories, matter at finite
temperature (using the BCS model of superconductivity as an illustrative
example), the Nambu--Jona-Lasinio model and the Schwinger model.  I conclude
with the outlook for further
development of this method.

\newpage
{\bf 1. ADVERTISEMENT}

Recently there has been a renewal of interest in analyzing quantum
chromodynamics starting from the action of the theory and using continuum
methods, rather than the lattice methods that have been extensively pursued.
Light-front methods have been emphasized as having the following
virtues:\\
\indent $\bullet$ Simple vacuum structure\\
\indent $\bullet$ Simple boosts\\
\indent $\bullet$ Intuitive wave function picture\\
\indent $\bullet$ Can be systematically improved\\
\indent $\bullet$ Nonperturbative\\
Light-front method have some drawbacks:\\
\indent $\bullet$ Not explicitly Lorentz invariant, so rotations are
complicated\\
\indent $\bullet$ Functions are required in counter terms for renormalization\\
\indent $\bullet$ The gauge is fixed, so gauge invariance is difficult to
check\\
\indent $\bullet$ Presence of zero modes\\
The amplitudes in the Haag expansion\cite{haag}
share the virtues of the light front method, and in
addition have the following virtues:\\
\indent $\bullet$ Have the same number of kinematic variables as Schr\"odinger
amplitudes\\
\indent $\bullet$ Obey three-dimensional equations that are
explicitly
covariant\\
\indent $\bullet$ In other words,
are as close to completely on-shell as possible in field
theory\\
\indent $\bullet$ Composite particles are treated in parallel with elementary
particles\\
\indent $\bullet$ Can be made crossing symmetric\\
This last property holds because, unlike the Tamm-Dancoff expansion,
the Haag expansion is made in the
fields, rather than in the states.
The Haag expansion method also has some drawbacks:\\
\indent $\bullet$ There are more graphs,
because many of the lines are on-shell\\
\indent $\bullet$ It is not clear in general how to truncate the expansion\\
\indent $\bullet$ In confining theories, a replacement must be found for
asymptotic fields\\
There are some hints how to do this from the Schwinger model.
If you like, you can use the Haag expansion on the light front and thus combine
the virtues and drawbacks of the two methods.

\newpage

{\bf 2. ASYMPTOTIC FIELDS AND THE HAAG EXPANSION}

Asymptotic fields have been part of quantum field theory at least
since the work of
Lehmann, Symanzik and Zimmermann\cite{lsz};
however, there are still misconceptions that
should be cleared up.  The asymptotic fields are free fields, because (at least
in theories which have neither massless particles nor confinement) the
particles
described by the asymptotic fields separate for large magnitude of the time.
The free field property of the asymptotic fields {\it does not depend on
unphysical ``adiabatic switching off'' of the interactions}.  The physical
assumption is that for $t \rightarrow \pm \infty$, particles either
(a) separate widely and thus move freely, since interactions fall off
exponentially in space
or
(b) stay close together and thus form a bound state that itself moves freely.
In this case asymptotic fields must be introduced for the bound state.
In either case, the exact eigenstates can be labelled by the quantum numbers of
free particles.  The limits for $t \rightarrow \pm \infty$ are the $out$ or
$in$
fields that make eigenstates at the corresponding limiting times.  The
asymptotic fields at finite times are the limiting fields brought back to
finite
times according to the free equations of motion.  The
unitary relation between these fields is given by the $S$-operator,
\beee
S \phi^{out} S^{\dagger} = \phi^{in}.
\eeee
We need asymptotic fields for those bound states that are stable in the
approximation under consideration; for example, in considering strong
interactions, pions would be taken to be stable and would receive asymptotic
fields.

The $in$ fields all have free commutation or anticommutation relations and free
equations of motion and commute or anticommute among themselves.  The same is
true for the $out$ fields.  Given the masses and spins of the fields, each
set, $in$
or $out$, is a completely known set of fields.  Thus each set is a convenient
set
of building blocks for the construction of solutions of quantum field theories.
The relation between the $in$ and the $out$ fields is nontrivial, given by the
S-operator.

The limits that define the asymptotic fields are subtle.  The relations that
appears in some books,
\beee
\phi(x) \rightarrow \phi^{out, ~in}(x), x^0 \rightarrow \pm \infty
\eeee
are ill-defined.
The proper limit is a weak operator limit that constructs an asymptotic field
of
a given mass $m$ from the neighborhood of the mass $m$ part of the relevant
(product of) (scalar) Lagrangian field(s)\cite{owg},
\beee
\phi^{out,~in}(x)=lim_{\tau \rightarrow \pm \infty}
[-\int_{{y^0}=\tau} \Delta(x-y;
m^2)\stackrel{\leftrightarrow}{\partial_{y^0}}\phi(y)d^3y].
\eeee
The relation to the mass $m$ part of the Lagrangian field is transparent from
the momentum space version,
\beee
\tilde{\phi}^{out,~ in}(k)\delta(k^2-m^2)=lim_{\tau \rightarrow \pm
\infty}\epsilon(k^0)\delta(k^2-m^2) \int dq^0 (k^0+q^0)\tilde{\phi}(q^0,{\bf
k})e^{-i(q^0-k^0)\tau}.
\eeee
For composite particles, one must use a product of the Lagrangian fields of the
elementary constituents.  More about that later.  When the field strength
renormalization diverges, which is generally the case in relativistic theories,
one must introduce an averaging over time in the definition of the limit.
See\cite{owg} for details of that.

To motivate the Haag expansion, recall that we expect that
a quantum field theory of particles in Hilbert
space has three complete and irreducible sets of field operators. Here complete
means that any state in the Hilbert space can be approximated by polynomials in
the smeared fields acting on a cyclic vector, usually the vacuum state, and
irreducible means that any operator that commutes with an irreducible set of
operators is a multiple of the identity.  The first such complete and
irreducible set is the set of Lagrangian fields, i.e., the fields that appear
in
the Lagrangian and in the action of the theory.  The second and third such sets
are the two sets of
asymptotic fields, including fields for bound states, if there are any.  Since
by themselves the set of $in$ fields are completely known, they are standard
building blocks from which the Lagrangian fields can be constructed.  The same,
of course, is true for the set of $out$ fields.  The Haag expansion is just the
expression of this idea.  For a theory with a single (scalar)
field and no bound states,
the Haag expansion is
\beee
A(x)=A^{in}(x) +\sum_{n=2}^{\infty}\int d^4x_1 \cdots d^4x_nf^{(n)}(x-x_1,
\cdots, x-x_n):A^{(in)}(x_1) \cdots A^{(in)}(x_n):,
\eeee
where the double dots indicate normal ordering of the $in$
fields\cite{haag,owg2}.
If there
are bound states, then $in$ fields for the bound states have to be introduced
where-ever the conservation laws of the theory allow\cite{gg}.
The physical vacuum is
the state annihilated by the positive frequency parts of the $in$ fields; only
the physical vacuum enters and it is a structureless state in this formulation.
The equations for the (scalar) $in$ fields are
\beee
(\Box + m^2)A^{(in)}(x)=0,
\eeee
\beee
[A^{(in)}(x), A^{(in)}(y)]_-=i \Delta(x-y;m^2),
\eeee
and the different $in$ fields completely commute or anticommute depending as
usual on whether the fields are bosons or fermions.  If spontaneous symmetry
breaking occurs, the Haag expansion starts with a c-no. term which is the
vacuum
matrix element of the scalar field\cite{ptp}.  Haag introduced this
expansion in 1955 to discuss questions of principle.  The present effort is
aimed a developing a practical calculational method based on his expansion.
The
$f^{(n)}$'s (``Haag'' amplitudes)
are multiple retarded commutator functions with all but one leg
on-shell\cite{glz}.  They automatically correspond to
connected graphs.  They obey the
Klein-Gordon equation (for scalar $in$ fields) in each argument.  Because of
this, the terms in the expansion can be replaced by
\beee
\int d^3x_i f^{(n)}(\cdots, x-x_i, \cdots)\cdots \stackrel{\leftrightarrow}
{\partial_{x_i^0}} \cdots : \cdots A^{(in)}(x_i)\cdots :
\eeee
which illustrates that the Haag amplitudes are {\it both} three-dimensional and
covariant.  The asymptotic limit applied to $A(x)$ gives $A^{(out)}(x)$ in
terms of $A^{(in)}(x)$.  The relation between the (anti)commutators of these
two,
\beee
[A^{(out)}(x), A^{(out)}(y)]_{\mp}=[A^{(in)}(x), A^{(in)}(y)]_{\mp}
\eeee
gives unitarity for all processes.  The equal-time commutation relations,
\beee
[A({\bf x}, t), A({\bf y}, t)]_-=0, [A({\bf x}, t),\dot{A}({\bf y}, t)]_-=
iZ_3^{-1}\delta({\bf x},{\bf y}),
\eeee
give generalizations of unitarity.  (The choice of $\dot{A}$ as the canonical
conjugate is valid for theories without derivative coupling.)  The Haag
amplitudes for bound states are like Schr\"odinger wave functions for the bound
states.  There are no relative-time coordinates as in the Bethe-Salpeter
amplitudes.  The Haag amplitudes for elastic scattering are the scattering
amplitudes with one leg off-shell and the other three legs on-shell.  The Haag
amplitudes for higher processes are related to scattering and production
amplitudes in a more complicated way.  This formalism is as close to
being completely on-shell as is possible in field theory.  Among the good
features of this approach is the fact that the Haag amplitudes have better
ultraviolet behavior than the totally off-shell time-ordered amplitudes.  This
is because the latter contain phase-space integrals that grow rapidly for large
numbers of particles in intermediate states.  The Haag amplitudes for the
four-dimensional derivative coupling model, which is exactly solvable,
illustrates this difference\cite{owg3}.
I emphasize that
there is nothing unorthodox about the Haag expansion; what is surprizing about
the Haag expansion is that, up to now,
it has not been developed into a powerful calculation
method.  That development is the goal of the present work.

{\bf 3. NONRELATIVISTIC FIELD THEORY WITH BOUND STATES}

\begin{flushleft}
{\bf 3a. The model}
\end{flushleft}

Consider a model with two spinless
nonrelativistic Fermi fields, $A(x)$ and $B(x)$,
with the Hamiltonian
\begin{eqnarray}
H & = & m_{A}\int\,d^{3}x\,A^{\dag}(x)\,A(x)+\frac{1}{2m_{A}}\int\,d^{3}x\,
\nabla_{{\bf x}}A^{\dag}(x)\cdot \nabla_{{\bf x}}A(x)
+(B\,~{\rm terms})\nonumber \\
  &   & \mbox{}
+\int_{x^0=y^0}\,d^{3}x\,d^{3}y\,
B^{\dag}(y)A^{\dag}(x)V_{AB}(|{\bf x}-{\bf y}|)A(x)B(y);
\end{eqnarray}
for simplicity I assumed an $AB$ interaction, but no $AA$ or $BB$
interaction.  The equation of motion for $A(x)$ is
\begin{equation}
{\em i}\partial_{x^{o}}A(x)
=(m_{A}-\frac{1}{2m_{A}}\nabla^{2}_{{\bf x}})A(x)
+\int_{x^0=y^0}\,d^{3}y\,B^{\dag}(y)V_{AB}(|{\bf x}-{\bf
y}|)B(y)A(x).\label{em}
\end{equation}
The asymptotic ($in$ or $out$) fields for (possibly composite) particles are
characterized by their rest energy $E$,
kinetic mass $m$ and spin $J$.  I suppress the spin in what follows.  The
kinetic mass is the mass that enters in the kinetic energy, ${\bf p}^2/2m$; for
composite particles, as discussed below, the kinetic mass is the sum of the
kinetic masses of the constituents, {\it without the binding energy}, because
of the Bargmann mass superselection rule described in Sec.\ 3b.  The
asymptotic fields obey the following free field equation and anticommutation or
commutation relations:
\begin{equation}
i\partial_{x^{o}}C^{in(out)}(x)
= (E-\frac{1}{2m}\nabla^{2})C^{in(out)}(x)
\end{equation}
\begin{equation}
[C^{in(out)}({\bf x},t),C^{\dag in(out)}({\bf y},t^{\prime})]_{\pm}
= {\cal D}({\bf x}-{\bf y},t-t^{\prime};E,m),
\end{equation}
\begin{equation}
{\cal D}({\bf x},t;E,m)=\frac{1}{2 \pi)^{3}}\int d^4k
\delta(k^0-E-\frac{{\bf k}^2}{2m}) e^{-ik^0t+i{\bf k}\cdot {\bf x}}.
\label{cal}
\end{equation}
Note that
\begin{equation}
{\cal D}({\bf x},0;E,m)=\delta({\bf x}),~~ \forall E, m.
\end{equation}

Using translation invariance, on can show that
the Haag expansion of the interacting field
$A(x)$ in terms of $in$ fields takes the
following form in position space (with an analogous expansion for the $B$
field)\cite{cg}
\begin{eqnarray}
A(x_{A})=A^{in} (x_{A}) + \sum_{i} \int d^{3}x_{B} d^{3}x_{i} f_{B;i}
(x_A-x_B;x_A-x_i)B^{\dag \, in} (x_{B})(ABi)^{in}
(x_{i})\nonumber\\
+ \int
d^3 x_B d^3 y_A d^3y_B f_{B;AB} (x_A-x_B;x_A-y_A,x_A-y_B)B^{\dag \, in} (x_B)
A^{in} (y_A) B^{in}(y_B)\nonumber\\
+\cdots,                                         \label{he}
\end{eqnarray}
where, since both the asymptotic fields and the Haag amplitudes obey free
equations, the integrals are independent of the times $x_B^0,x_i^0,
y_A^0$ and $y_B^0$ because of the translation
invariance of the Schr\"odinger scalar products.
Label the Haag amplitude that is
the coefficient of a product of (asymptotic) creation and annihilation
operators
by the labels of the operators.  Labels to the left of the semicolon
are creation
operators, to the right are annihilation operators; the two-body
$(AB)$ bound state in
level $i$ is labeled by $i$.

Some calculations are simpler in momentum space, therefore define
\begin{equation}
A(x) =\int\,d^{4}k\,e^{-{\em i}k\cdot x}\tilde{A}(k),
\end{equation}
\begin{equation}
V_{AB}(|{\bf x}-{\bf y}|)
= \frac{1}{(2\pi)^3}\int\,d^{3}q\,e^{{\em i}{\bf q}\cdot
({\bf x}-{\bf y})}\tilde{V}_{AB}({\bf q}).
\end{equation}
{}From now on, drop the tildes on $\tilde{A}$ and $\tilde{V}$.
Transforming the equation of motion to momentum space yields
\begin{eqnarray}
(k^{0}_A-m_{A}-\frac{{\bf k}_A^2}{2m_{A}})A(k_A)=
\int\,d^{4}k_Bd^{4}p_Bd^{4}p_A
\delta(k_A+k_{B}-p_{B}-p_{A})\nonumber \\
\times B^{\dag}(k_B)V_{AB}({\bf p}_B-{\bf k}_B)B(p_B)A(p_A).
\end{eqnarray}
Define
\begin{equation}
C^{in}(x)=(2 \pi)^{-3/2}\int d^4k \delta(k^0-E_C-{\bf k}^2/2m_C)
e^{-ik^0t+i{\bf k}\cdot {\bf x}}c^{in}({\bf k}),
\end{equation}
\begin{equation}
[c^{in}({\bf k}),c^{\dagger~in}({\bf l})]_+=\delta({\bf k}-{\bf l})  \label{ft}
\end{equation}
\begin{equation}
f_{B;i}(x; y)=\frac{1}{(2\pi)^3}\int d^3k_1 d^3k_2
e^{i(m_B+\frac{{\bf k}_1^2}{2m_B})x^0-i{\bf k}_1\cdot{\bf x}
-i(m_{AB}-\epsilon_i+\frac{{\bf k}_2^2}{2m_{AB}})y^0+i{\bf k}_2\cdot {\bf y}}
\tilde{f}_{B;i}({\bf k}_1,{\bf k}_2)
\end{equation}
and similar definitions for other Fourier transforms chosen so
that powers of
$2\pi$ are absent from most of the momentum-space formulas.  The result is
\[A(k_A)=(2 \pi)^{-3/2}:k_{A}:\delta(k^0_A-m_A-\frac{{\bf k}_A^2}{2m_A})\]
\[+\int\,d^{3}k_{B}d^{3}k_{i}\,
\delta(k^0_A+\frac{{\bf k}^2_{B}}{2m_B}-m_A+\epsilon_i-
\frac{{\bf k}^2_{i}}{2m_{AB}})\delta({\bf k}_A+{\bf k}_{B}-{\bf k}_{i})
\tilde{f}_{B;i}({\bf k}_B;{\bf k}_{i})
:k_B^{\dag}k_{i}:\]
\[+\int\,d^{3}k_{B}d^{3}p_{B}d^{3}p_{A}\,
\delta(k_A^0+\frac{{\bf k}_{B}^2}{2m_B}-m_A-\frac{{\bf p}^{2}_{A}}{2m_A}
-\frac{{\bf p}^{2}_{B}}{2m_B})\delta({\bf k}_A+{\bf k}_{B}
-{\bf p}_{A}-{\bf p}_{B})\]
\begin{equation}
\times\tilde{f}_{B;AB}({\bf k}_B;{\bf p}_A,{\bf p}_B)
:k^{\dag}_{B}p_{A}p_{B}:+\cdots,
\end{equation}
where I define
$:p_{A}:\equiv a^{in}(p)$, with normalization as given in
Eq.(\ref{ft}), etc.  Note
that I am expanding in terms of $in$ fields;
there are analogous expansions in terms
of $out$ fields.  In the next section I derive the constraints on the $f$'s
that
follow from Galilean invariance.

\begin{flushleft}
{ \bf 3b. Galilean Invariance}
\end{flushleft}

Bargmann\cite{barg} showed that the unitary projective representations
(i.e., representations up to a factor) of the
Galilean group that occur in the quantum mechanics of nonrelativistic particles
cannot be reduced to vector (i.e., true) representations.  This contrasts with
the corresponding situation for the Poincar\'e and Lorentz groups
(and indeed most other physically interesting groups), where the
representations
can be reduced to true representations.  The explicit mass
parameter in the phases leads to the Bargmann superselection rule that the
sum of the masses (that appear in the kinetic terms) must be conserved in every
process.  Nonetheless, bound states can be formed and particles can be created
and annihilated, provided the Bargmann superselection rule is obeyed.

Note that, for example, a bound state of particles of masses $m_A$ and $m_B$
with binding energy $\epsilon$ has energy $E=m_A+m_B-\epsilon+
{\bf k}^2/2 m_{AB}$, rather than $E=m_A+m_B-\epsilon+
{\bf k}^2/2 (m_{AB}-\epsilon)$ as one might expect from the
nonrelativistic limit of a relativistic bound state with rest energy
$m_A+m_B-\epsilon$.  (I use the abbreviation $m_{AB}=m_A+m_B$.)
Another manifestation of this effect is that for this
same bound state the momentum transforms under Galilean boosts as
${\bf k} \rightarrow {\bf k}+m_{AB}{\bf v}$, rather than as
${\bf k} \rightarrow {\bf k}+(m_{AB}-\epsilon){\bf v}$.

If the
projective representation has the form
\begin{equation}
U(G_2)U(G_1)=\omega(G_2,G_1)U(G_2G_1)
\end{equation}
then another projective representation is equivalent to this if the other
representation has the factor system $\omega^{\prime}(G_2,
G_1)=[\phi(G_2)\phi(G_1)/\phi(G_2G_1)]\omega(G_2,G_1)$, where $\phi$ has
modulus
one.  This arbitrariness allows simplification of some formulas.

Bargmann gives as the Galilean transformation of a
nonrelativistic scalar wave function,
\begin{equation}
(T(G)\psi)(x)=e^{-i\theta(G,x)}\psi(G^{-1}x),
\end{equation}
where $x=({\bf x},t)$, the Galilean transformation is
$Gx=(R{\bf x}+{\bf v}t+{\bf a}, t+b)$, $G=({\bf a}, b, R,{\bf v})$,
where ${\bf a}$ and
$b$ are space and time translations, $R$ is a rotation and ${\bf v}$ is a
boost, and
$\theta(G,x)=m(\frac{1}{2}{\bf v}^2t-{\bf v}\cdot {\bf x})$.
To infer the corresponding transformation for a nonrelativistic scalar
field, I require
\begin{equation}
U(G)A(\psi)U^{\dagger}(G)=A(\psi_G),~~\psi_G(x)=(T(G)\psi)(x)=e^{-i\theta(G,x)}
\psi(G^{-1}x),
\end{equation}
\begin{equation}
A(\psi)=\int A(x)\psi(x)d^4x.
\end{equation}
Then
\begin{equation}
U(G)A(x)U^{\dagger}(G)=e^{-i\theta_A(G,Gx)}A(Gx),~~\theta_A(G,Gx)=m_A
[\frac{1}{2}{\bf v}^2(t+b)-{\bf v}\cdot (R{\bf x}+{\bf v}t+{\bf a})].\label{f}
\end{equation}
If the field has spin $s$, then $A$ on the left hand side is replaced by $A_i$
and $A$ on the right hand side is replaced by $\sum_j A_j D^{(s)}_{ji}(G)$,
where $D^{(s)}$ is a representation of $SU(2)$, which is the little group in
this case.
The corresponding transformation holds for $B$ with $m_B$ replacing $m_A$.
Asymptotic fields transform the same way.    The
implications of the transformation law for the Haag amplitudes
is found by transforming the interacting field in two ways: (1) insert the
right-hand side of Eq.(\ref{f}) in the Haag expansion, or
(2) transform each of the
asymptotic fields and then change variables to get the transformation into the
amplitudes.  The two amplitudes $f_{B;i}$ and $f_{B;AB}$ obey
\[f_{B;i}(G(x_A-x_B);G(x_A-x_i))=\]
\begin{equation}
 e^{i\theta_A(G,Gx_A)+i\theta_B(G,Gx_B)-i\theta_{AB}(G,Gx_i)}
f_{B;i}(x_A-x_B;x_A-x_i),
\end{equation}
\[f_{B;AB}(G(x_A-x_B);G(x_A-y_A),(G(x_A-y_B))=\]
\begin{equation}
e^{i\theta_A(G,Gx_A)+i\theta_B(G,Gx_B)
-i\theta_{A}(G,Gy_A)-i\theta_B(G,Gy_B)}f_{B;AB}(x_A-x_B;x_A-y_A,x_A-y_B).
\end{equation}
Note that $\theta_{AB}$ is independent of the bound state $i$ because of the
Bargmann mass
superselection rule.  The combination of phases in the first of these is
\[\theta_A(G,Gx_A)+\theta_B(G,Gx_B)-\theta_{AB}(G,Gx_i)=\]
\begin{equation}
-\frac{1}{2}{\bf v}^2(m_Ax_A^0+m_Bx_B^0-m_{AB}x_i^0)
-{\bf v}\cdot R(m_Ax_A+m_Bx_B-m_{AB}x_i).
\end{equation}
The transformation law is {\it not} satisfied by having a delta function in
the space and time coordinates identifying the coordinate $x_i$ with the
center-of-mass of particles $A$ and $B$, although at equal times such a delta
function does occur for the space coordinates.  The way in which the
transformation laws are satisfied is best seen in momentum space, to which I
now turn.

The corresponding transformations in momentum space are
\begin{equation}
(V(G)\phi)(k)=
e^{-i\Omega(G,k)}\phi(G^{-1}k),
\end{equation}
\begin{equation}
\Omega(G, k)=({\bf k}-m{\bf v})\cdot{\bf a}-
(k^o-\frac{1}{2}m{\bf v}^2)b,
\end{equation}
where $k=({\bf k}, E)$,~~$Gk=(R{\bf k}+m{\bf v}, E+{\bf v}\cdot R{\bf k}+
\frac{1}{2} m {\bf v}^2)$, and
$G^{-1}k=(R^{-1}({\bf k}-m{\bf v}),
E-{\bf k}\cdot {\bf v}+\frac{1}{2}m{\bf v}^2)$.
The momentum space transformation law for the field is induced in parallel with
the derivation of the position space law.  The result is
\begin{equation}
W(G)A(k)W^{\dagger}(G)=e^{-i\Omega_A(G,-Gk)}A(Gk),
\end{equation}
where $\Omega_A(G,-Gk)=(E+{\bf v}\cdot R{\bf k})b-R{\bf k}\cdot {\bf a}$.
In the transformation law for the Haag amplitudes, all the phase factors
cancel and the result for--say--the second term in
the Haag expansion is what one would expect naively,
\begin{equation}
\tilde{f}_{B;i}({\bf k}_B;{\bf k}_i)=
\tilde{f}_{B;i}(R({\bf k}_B-m_B{\bf v});R({\bf k}_i-m_{AB}{\bf v})).
\end{equation}
Thus I can choose the ${\bf v}={\bf k}_i/m_{AB}$ so that the bound-state
momentum vanishes and eliminate the second argument of $f_{B;i}$,
\begin{equation}
\tilde{f}_{B;i}({\bf k}_B;{\bf k}_{i})=
\tilde{f}_{B;i}({\bf k}_B-\frac{m_B}{m_{AB}}{\bf k}_i,{\bf 0})
\equiv \tilde{F}_{B;i}({\bf k}_B-\frac{m_B}{m_{AB}}{\bf k}_i). \label{gi}
\end{equation}
For the spinless case,
$\tilde{F}_{B;i}({\bf k})=\tilde{F}_{B;i}(R{\bf k}).$
All these results are exact, valid in any Galilean frame.  The
extension to fields
with spin is straightforward.  It is worth noting that the Poincar\'e
transformation law in a relativistic theory is simpler than the Galilean
transformation law because the Bargmann phase is absent for the
Poincar\'e group.

Taking account of Galilean invariance, one finds that
the position-space Haag amplitude is
\begin{eqnarray}
f_{B;i} (x;y)& = & (2\pi)^{-3} \int d^{3}k  d^{3}k_{i}
exp[i (m_{B} + \frac{1}{2m_{B}}({\bf k} +
\frac{m_{B}}{m_{AB}}{\bf k}_{i})^{2} x^0 - i ({\bf k} +
\frac{m_{B}}{m_{AB}}   {\bf k}_{i}) \cdot {\bf x}] \nonumber \\
& &\times exp[-i (m_{AB} - \epsilon_i + \frac{{\bf k}^2_{i}}{2m_{AB}})
y^0 + i {\bf k}_{i} \cdot {\bf y}] \tilde{f}_{B;i} ({\bf k};{\bf 0}).
\end{eqnarray}
The integral over ${\bf k}_i$ can be done, but the result is complicated and
not useful, except when all times are equal, in which case the result is both
simple and useful,
\begin{equation}
f_{B;i}({\bf x}_A-{\bf x}_B;{\bf x}_A-{\bf x}_i)=
\delta({\bf x}_{i}-\frac{m_A {\bf x}_A+m_B {\bf x}_B}{m_{AB}})
F_{B;i}({\bf x}_A-{\bf x}_B),               \label{29}
\end{equation}
\begin{equation}
F_{B;i}({\bf x})=\int d^3k e^{-i {\bf k}\cdot {\bf x}} f_{B;i}({\bf k};{\bf
0}).
                 \label{et}
\end{equation}
Using the constraints due to Galilean invariance, the Haag expansion in
$x$-space at equal times takes the form
\[ A({\bf x}_A)=A^{in}({\bf x}_A)+\sum_i\int F_{B;i}({\bf x}_A-{\bf x}_B)
B^{\dagger in}({\bf x}_B)(ABi)^{in}
(\frac{m_A{\bf x}_A+m_B{\bf x}_B}{m_{AB}})d^3x_B \]
\[ +\int d^3r^{\prime} d^3r F_{B;AB} ({\bf r^{\prime}};{\bf r})
B^{\dagger in}({\bf x}_A - {\bf r}^{\prime}) A^{in}
({\bf x}_A + \frac {m_B ({\bf r} - {\bf r}^{\prime})}{m_{AB}})
B^{in} ({\bf x}_A - \frac{m_A {\bf r} + m_B{\bf r}^{\prime}}{m_{AB}}) \]
\begin{equation}
+\cdots.
\end{equation}
In momentum space, the expansion is
\[ A(k_A)=\frac{1}{(2 \pi)^{3/2}}
:k_A:\delta(k_A^0-m_A-\frac{{\bf k}_A^2}{2m_A})\]
\[+\int d^3k_B
\delta(k_A^0+\frac{{\bf k}_B^2}{2m_B}-m_A+\epsilon_i-
\frac{({\bf k}_A+{\bf k}_B)^2}{2m_{AB}})\tilde{F}_{B;i}(\frac{m_A{\bf k}_B
-m_B{\bf k}_A}{m_{AB}}):k^{\dagger}_B({\bf k}_A+{\bf k}_B)_i:  \]
\[+\int d^3k_B d^3p_B d^3p_A\delta(k_A^0+\frac{{\bf k}_B^2}{2m_B}-m_A-
\frac{{\bf p}_A^2}{2m_A}-\frac{{\bf p}_B^2}{2m_B})\delta({\bf k}_A+{\bf k}_B
-{\bf p}_A-{\bf p}_B)\]
\begin{equation}
\times \tilde{F}_{B;AB}(\frac{m_A{\bf k}_B-m_B{\bf k}_A}{m_{AB}};
\frac{m_A{\bf p}_B-m_B{\bf p}_A}{m_{AB}}):k^{\dagger}_Bp_Ap_B: +\cdots .
\end{equation}

\begin{flushleft}
{\bf 3c. Two-Body Bound State}
\end{flushleft}

To derive the equation for the two-body bound state, insert the Haag
expansion Eq.(\ref{he}) in the equation of motion Eq.(\ref{em}),
renormal order, and equate the
coefficients of the terms with the operators $B^{\dagger in}(AB_i)^{in}$.
After
commuting or anticommuting with the relevant $in$ fields, the result is
\begin{equation}
(i\frac{\partial}{\partial x^0_A} - m_A +\frac{1}{2m_A}{\bf \nabla}^2_{x_A}
-V({\bf x}_A-{\bf x}_B))f_{B;i}(x_A-x_B;x_A-x_i)=0
\end{equation}
It is convenient to eliminate the time derivative by using
\mbox{$\partial/\partial x^0_A=-\partial/\partial x^0_B
-\partial/\partial x^0_i$,} the independence of the
Schr\"odinger scalar product on the time and the free equations satisfied by
the
$in$ fields
to find free equations for the $x_B^0$ and $x_i^0$ dependences of $f_{B;i}$.
The results are
\begin{equation}
(i\frac{\partial}{\partial x^0_B}-m_B+\frac{1}{2m_B}{\bf \nabla}^2_{x_B})
f_{B;i}=0,
\end{equation}
\begin{equation}
(i\frac{\partial}{\partial x^0_i}+m_{AB}-\epsilon_i-
\frac{1}{2m_{AB}}{\bf \nabla}^2_{x_i})f_{B;i}=0.
\end{equation}
The equation without time derivatives is
\begin{equation}
[-\frac{1}{2m_A}{\bf \nabla}^2_{x_A}-\frac{1}{2m_B}{\bf \nabla}^2_{x_B}+
V({\bf x}_A-{\bf x}_B)]f_{B;i}=(\epsilon_i
-\frac{1}{2m_{AB}}{\bf \nabla}^2_{x_i})f_{B;i}.
\end{equation}
Now using Eq.(\ref{29}) the usual Schr\"odinger equation for
$F_{B;i}$
results,
\begin{equation}
[-\frac{1}{2\mu}{\bf \nabla}^2_{r_{AB}}+
V({\bf r}_{AB})]F_{B;i}=-\epsilon_iF_{B;i},~~\frac{1}{\mu}=\frac{1}{m_A}+
\frac{1}{m_B},
\end{equation}
where the reduced mass enters.
This establishes that $F_{B;i}$ is the Schr\"odinger wave function of the bound
state.
Note that the bound-state amplitude is given
{\it exactly} in any reference frame in terms of the amplitude in the rest
frame
of the bound state.  (The corresponding statement also holds for other
amplitudes, as well as for relativistic theories.)

\begin{flushleft}
{\bf 3d.  Two-Body Scattering}
\end{flushleft}

Two-body scattering is described in position space at equal times by the
amplitude
\begin{equation}
f_{B;AB} ({\bf x}_A - {\bf x}_B,0;
{\bf x}_A - {\bf y}_A,0, {\bf x}_B - {\bf y}_B,0) = F_{B;AB}
({\bf x}_A - {\bf x}_{B}; {\bf y}_A - {\bf y}_B) \delta
({\bf R^{\prime}} - {\bf R}),
\end{equation}
\begin{equation}
F_{B;AB} ({\bf x}; {\bf y}) = (2 \pi)^{-3/2}
\int d^3k^{\prime}d^3k \tilde{f}_{B;AB}({\bf k}^{\prime}; -{\bf k},
{\bf k}) \exp[i(-{\bf k}^{\prime}\cdot ({\bf x}_A - {\bf x}_B) +
{\bf k}\cdot({\bf y}_A - {\bf y}_B)],
\end{equation}
$${\bf R}^{\prime} = \frac{m_A {\bf x}_A + m_B{\bf x}_B}
{m_{AB}},~~ \hspace{.2in} {\bf R} =
\frac{m_A{\bf y}_A + m_B{\bf y}_B}{m_{AB}}.$$
I prefer to discuss two-body scattering in momentum space, using
the amplitude
$\tilde{f}_{B;AB}({\bf k}_{B};{\bf p}_{A},{\bf p}_{B})$ which is the
coefficient of the term $:k^{\dag}_{B}p_{A}p_{B}:$ in the Haag expansion of
$A(k)$.  The procedure for finding the equation for
$\tilde{f}_{B;AB}$ is analogous
to that for the two-body bound state amplitude.  One finds
\begin{eqnarray}
(\frac{{\bf p}_{A}^{2}-({\bf p}_A+{\bf p}_B-{\bf k}_{B})^{2}}{2m_{A}}
+\frac{{\bf p}_{B}^{2}-{\bf k}_{B}^{2}}{2m_{B}})
\tilde{f}_{B;AB}({\bf k}_{B};{\bf p}_{A},{\bf p}_{B}) &=&\nonumber \\
V_{AB}(|{\bf k}_{B}-{\bf p}_{B}|)
+\int\,d^{3}k_B^{\prime}\,V_{AB}(|{\bf k}_B-{\bf k}_B^{\prime}|)
\tilde{f}_{B;AB}({\bf k}_B^{\prime};{\bf p}_{A},{\bf p}_B).
\end{eqnarray}
Galilean invariance relates $\tilde{f}_{B;AB}$ at arbitrary
momenta to itself in the center-of-mass,
\begin{equation}
\tilde{f}_{B;AB}({\bf k}_{B};{\bf p}_{A},{\bf p}_{B}) =
\tilde{f}_{B;AB}(R({\bf k}_{B}-m_B{\bf v});R({\bf p}_{A}-m_A{\bf v}),
R({\bf p}_{B}-m_B{\bf v})).
\end{equation}
By choosing ${\bf v}=({\bf p}_A+{\bf p}_B)/m_{AB}$,
I can replace $\tilde{f}_{B;AB}$ by a function of one fewer variable,
\begin{equation}
\tilde{f}_{B;AB}({\bf k}_{B};{\bf p}_{A},{\bf p}_{B})=
\tilde{F}_{B;AB}({\bf k};{\bf p}),   \label{42}
\end{equation}
where here and below,
${\bf k}=(m_A{\bf k}_B-m_B{\bf k}_A)/m_{AB}$,
${\bf p}=(m_A{\bf p}_B-m_B{\bf p}_A)/m_{AB}$ and I used conservation of
momentum to introduce ${\bf k}_A$.  The momenta ${\bf p}$
and ${\bf k}$ are the center-of-mass momenta of particle $B$ in the initial and
the final state, respectively.
The elastic scattering equation becomes
\begin{equation}
\frac{1}{2\mu}({\bf p}^2-{\bf k}^2)
\tilde{F}_{B;AB}({\bf k};{\bf p})
=V(|{\bf k}-{\bf p}|)+\int d^3k^{\prime}
V({\bf k}-{\bf k}^{\prime})\tilde{F}_{B;AB}
({\bf k}^{\prime};{\bf p}),
\end{equation}
The solution is the Born series,
\begin{eqnarray*}
\lefteqn{\tilde{F}_{B;AB}({\bf k};{\bf p})=} \nonumber \\
& & \tilde{G}_R({\bf k};{\bf p})V(|{\bf k}-{\bf p}|)+
\tilde{G}_R({\bf k};{\bf p})\int d^3k^{\prime}
V(|{\bf k}-{\bf k^{\prime}}|)\tilde{G}_R({\bf k^{\prime}};{\bf p})
V(|{\bf k^{\prime}}-{\bf p}|)+ \cdots,
\end{eqnarray*}
where $\tilde{G}_R({\bf k};{\bf p})=
[({\bf p}^2-{\bf k}^2)/2 \mu-i\epsilon]^{-1}$.

The amplitude $\tilde{F}_{B;AB}$
is closely related to the $T$-matrix element for $AB$
scattering. The $S$-matrix element is
\begin{equation}
S(k_{A}, k_{B};p_{A}, p_{B}) \equiv
_{out}\langle k_{B},k_{A}|p_{A},p_{B}\rangle_{in}\equiv
\langle 0|:k_{B}^{out}k_{A}^{out}p_{A}^{\dag}p_{B}^{\dag}:|0\rangle,
\end{equation}
where I remind the reader that $:k_A:$, etc.,
stands for the $in$ field.
In order to eliminate the $out$ fields in terms of the $in$ fields, use
the definitions,
\begin{equation}
A^{in(out)}(x)=\lim_{\tau \rightarrow -\infty(\infty)}\int_{y^{0}=\tau}\,
d^{3}y\,{\cal D}(x-y;m_{A},m_{A})A(y),
\end{equation}
where ${\cal D}$ was defined in Eq.(\ref{cal}).
The nonrelativistic analog of the reduction formula follows from calculating
$\int d^4y \partial/ \partial_{y^0}{\cal D}(x-y;m_A,m_A)A(y)$ in two ways:
performing the integral and carrying out the derivative.  The result\cite{owg}
is
\begin{equation}
A^{out}(x)-A^{in}(x)=\int\,d^{4}y\,{\cal D}(x-y;m_{A},m_A)
(\partial_{y^0}+i m_{A}-
\frac{i}{2m_{A}}\nabla^{2}_{{\bf y}})A(y).
\end{equation}
Fourier transforming this one gets, after removing a factor of
$\delta(k^{o}-m_{A}-{\bf k}^{2}/2m_{A})$,
\begin{equation}
\frac{1}{(2 \pi)^{3/2}}(a^{out}({\bf k})-a^{in}({\bf k}))=-2 \pi i
(k^{o}-m_{A}-\frac{{\bf k}^{2}}{2m_{A}})A(k).  \label{in-out}
\end{equation}
The right-hand-side is non-vanishing (and there is
scattering) only when $A(k)$ has a pole at
$(k^{o}-m_{A}-{\bf k}^{2}/2m_{A})=0$.  Since
$a^{\dagger~out}(k)|0\rangle=a^{\dagger~in}(k)|0\rangle$ for stable particles,
the only $out$ operator in the $S$-matrix element
$\langle 0| :k^{out}_B::k^{out}_A::p^{~\dagger in}_A::p^{~\dagger in}_B:|0
\rangle$ that must be eliminated using Eq.(\ref{in-out}) is
$:k_{A}^{out}:$.  The result is
\[S(k_{A}, k_{B};p_{A}, p_{B})  =  \delta({\bf k}_A-{\bf p}_A)
\delta({\bf k}_B-{\bf p}_B)
-2 \pi i
\delta(\frac{k_A^2}{2m_A}+\frac{{\bf k}^2_B}{2m_B}-\frac{{\bf p}^2_A}{2m_A}
-\frac{{\bf p}^2_B}{2m_B}) \]
\beee
\times \delta({\bf k}_A+{\bf k}_B-{\bf p}_A-{\bf p}_B)
(\frac{k_A^2}{2m_A}+\frac{{\bf k}^2_B}{2m_B}-\frac{{\bf p}^2_A}{2m_A}\\
-\frac{{\bf p}^2_B}{2m_B})\tilde{F}_{B;AB}({\bf k};{\bf p}),
\eeee
where again ${\bf k}$ and ${\bf p}$ are defined below Eq.(\ref{42}).
Thus the reduced $T$-matrix
 for elastic scattering on the momentum shell is
\begin{equation}
t({\bf k}_A,{\bf k}_B;{\bf p}_A,{\bf p}_B)=
[\frac{{\bf p}_A^2}{2m_A}
+\frac{{\bf p}_B^2}{2m_B}-\frac{{\bf k}_A^2}{2m_A}-
\frac{{\bf k}_B^2}{2m_B}]
\tilde{F}_{B;AB}({\bf k};{\bf p}).
\end{equation}
I emphasize that because the Haag amplitude is the scattering amplitude with
one leg off shell, it contains the information necessary for calculations in
the three-body sector. This contrasts with the on-shell scattering amplitude,
which does not suffice for such calculations.

\begin{flushleft}
{\bf 3e. Anticommutation Relations}
\end{flushleft}

In this section I
show that the canonical (equal time)
anticommutation relations of the Lagrangian fields imply general
relations among Haag
amplitudes, independent of the equations of motion of the specific theory.
For example, the vanishing of the canonical anticommutator $[A,B]_+$ at equal
times,
considered for the coefficient of the bound state $in$ field for state $i$,
gives
\begin{equation}
F_{A;i}({\bf y}-{\bf x})=F_{B;i}({\bf x}-{\bf y})\equiv F_i({\bf x}-{\bf y})
\end{equation}
where I took $(ABi)^{in}({\bf R})=-(BAi)^{in}({\bf R})$ because of the Fermi
statistics of $A$ and $B$.
Thus the apparent asymmetry in the treatment of the
constituents of the bound state, due to the fact that the Haag amplitude that
serves as the two-body wave
function of the $(AB)$ bound state in the Haag expansion of the $A$ field has
the $A$ particle off-shell and the $B$
particle on-shell, while these roles are interchanged for the amplitude for the
same bound state in the Haag expansion of the $B$ field, is not a real
asymmetry.  These two amplitudes determine each other uniquely.  The analogous
result for the off-shell elastic scattering amplitudes is
\begin{equation}
F_{B;AB} ({\bf x}-{\bf y};{\bf r})=F_{A;BA} ({\bf y}-{\bf x};-{\bf r}) \equiv
F_{AB} ({\bf x}-{\bf y};{\bf r}).
\end{equation}
Again the two apparently different off-shell amplitudes uniquely determine each
other.

The consequence for elastic scattering is
\begin{eqnarray}
\lefteqn{t({\bf k}_A,{\bf k}_B;{\bf p}_A,{\bf p}_B)-
(t({\bf p}_A,{\bf p}_B;{\bf k}_A,{\bf k}_B))^{\ast}=}  \nonumber \\
& & (2 \pi)^{5/2} \int d^3q_A d^3q_B \delta(\frac{{\bf k}_A^2}{2m_A}
+\frac{{\bf k}^2_B}{2m_B}-\frac{{\bf q}^2_A}{2m_A}
-\frac{{\bf q}_B^2}{2m_B})\delta({\bf k}_A+{\bf k}_B-{\bf q}_A-{\bf q}_B)
\nonumber  \\
& & \times t({\bf k}_A,{\bf k}_B;{\bf q}_A,{\bf q}_B)
(t({\bf p}_A,{\bf p}_B;{\bf q}_A,{\bf q}_B))^{\ast},
\label{elun}
\end{eqnarray}
where ${\bf k}$ and ${\bf p}$ are as defined below Eq.(\ref{42}).  This
is elastic unitarity.

The
canonical anticommutator $[A,A^{\dagger}]_+$ at equal times leads to a
generalization of unitarity,
\[ \frac{1}{(2 \pi)^{3/2}}
(\tilde{F}_{B;AB}({\bf k};{\bf p})+\tilde{F}_{B;AB}^{\ast}({\bf p};{\bf k}))\]
\begin{equation}
=\sum_i\tilde{F}_{B;i}({\bf k})\tilde{F}^{\ast}_{B;i}({\bf p})
+\int d^3q\tilde{F}_{B;AB}({\bf k};{\bf q})
\tilde{F}^{\ast}_{B;AB}({\bf p};{\bf q}),
\end{equation}
where again ${\bf k}$ and ${\bf p}$ are as defined below Eq.(\ref{42}) and I
have used momentum conservation, ${\bf k}_A+{\bf k}_B={\bf p}_A+{\bf p}_B$.
By taking the
appropriate limit, I recover the elastic unitarity relation, Eq.(\ref{elun}).
On taking into account the
relations between the Haag amplitudes in the expansions
of $A$ and of $B$, one find that these are all the independent
two-body relations obtained from the
anticommutation relations.

There are also quadratic relations between the amplitudes for the
$(ABi)$ and $(ABj)$ bound states and the amplitudes for the breakup of these
bound states due to scattering with the $A$ or $B$ particle.
Since this involves
a higher sector, I do not give this relation here.

\begin{flushleft}
{\bf 3f. Construction of the asymptotic field for the bound state}
\end{flushleft}

In this section I show how to construct the asymptotic field for the bound
state from a product of Lagrangian fields.  My suggestion differs from that
proposed by Nishijima\cite{nish} and by Zimmermann\cite{zimm}.
The procedure is to multiply the appropriate Lagrangian fields at separated
space points, integrate
with the bound-state amplitude in the relative coordinate,
and take the asymptotic limit.  If the $in$ field expansions of the Lagrangian
fields are inserted and the resulting expression normal ordered, then the
$t \rightarrow -\infty $ limit gives the $in$ field bound state operator and
the
$t \rightarrow \infty $ limit gives a reduction formula for
the $out$ field bound state operator.  The result is
\[(ABi)^{in(out)}(x)=\lim_{\tau \rightarrow -\infty(\infty)}\int_{y^{0}=\tau}\,
d^{3}y\,{\cal D}(x-y;m_{AB}-\epsilon_i,m_{AB})F^{\ast}(w)\]
\begin{equation}
\times \frac{1}{2} [B(y-\frac{m_A}{m_{AB}}w),A(y+\frac{m_B}{m_{AB}}w)]_-d^3w.
\end{equation}
A straightforward calculation shows this limit is $(ABi)^{in}(x)$ for $\tau
\rightarrow -\infty$ and the leading term for $\tau
\rightarrow \infty$ is $(ABi)^{out}(x)$.  Both results are what we expect.  The
later terms in the Haag expansion for $(ABi)^{out}(x)$ are in a higher sector
that I don't discuss here.

I derived many results of the nonrelativistic quantum mechanics of
two-particle systems in a unified way with particular attention to Galilean
invariance,
taking into account the fact that the representations of the Galilean
group in quantum mechanics are necessarily representations up to a factor,
rather than vector representations.  The Haag amplitude for the simplest term
with the
two-body bound-state operator is precisely the Schr\"odinger wave function of
the two-body bound state.  The amplitude for the term with three {\it in}
fields is the scattering amplitude with one leg off-shell.
These interpretations carry over to explicitly
covariant relativistic theories,
where the corresponding Haag amplitude is defined on
three-dimensional manifolds, but is covariant.  Of course in the relativistic
case, a bound state that is mainly a two-body state also will have amplitudes
to
be composed of more than two particles.

{\bf 4. The NAMBU--JONA-LASINIO MODEL}

The Haag expansion
is effective in treating the Nambu--Jona-Lasinio model in one-loop
approximation. In particular,
the Haag expansion sums crossed as well as direct graphs\cite{gm,orr}, in
contrast to the usual methods which sum only direct graphs.
Further, using the Haag expansion one deals directly with the bound-state
wavefunction (or
amplitude); one does not have to extract the bound-state amplitude as the
residue of a pole in the scattering amplitude.

The Lagrangian of
the model without isospin is
\beee
{\cal L}=i\bar{\psi}\! \partial \!\!\! / \psi
-\frac{1}{2}g_0[(\bar{\psi}\gamma_{\mu}\psi)\gamma^{\mu}\psi
-(\bar{\psi}\gamma_{\mu}\gamma_5\psi)\gamma^{\mu}\gamma_5\psi].
\eeee
To take account of operator ordering, symmetrize or antisymmetrize the
operator products.  After going to momentum space, the equation of motion is
\begin{eqnarray}
q \!\!\! /  \psi(q)&=&-\frac{1}{2}g_0
\int d^4p_1 d^4p_2 d^4p_3 \delta(q-p_1-p_2-p_3)\\
& &\{[[\bar{\psi}(p_1),\psi(p_2)]_-,\psi(p_3)]_+
-[[\bar{\psi}(p_1),\gamma_5\psi(p_2)]_-,\gamma_5\psi(p_3)]_+\}
\end{eqnarray}
The lowest approximation, to take $\psi(p)=\psi^{in}(p)\delta(p^2-m^2)$, leads
to the gap equation,
\beee
\mu=\frac{4g_0 \mu}{(2 \pi)^3} \int d^4p \delta(p^2-\mu^2).
\eeee
Surely, this is a simple derivation of this result!  Since this model is
nonrenormalizable, one must cut off the integral.  This can be done
covariantly,
if desired.  For $0 < \pi^2/g_0 \Lambda^2 <1$, where $\Lambda$ is the cutoff,
there are three solutions: $\mu=0$ and $\mu=\pm
m$.  The first is the unbroken symmetry solution; the last two are equivalent
broken symmetry solutions.  To decide which solution is the stable one,
calculate the matrix element of the Hamiltonian in the corresponding
vacuum state.  The result is that the broken symmetry solutions have lower
vacuum matrix elements and are thus the correct solutions.

To find the bound states, consider the term in the Haag expansion with the
product $:\psi^{in} B^{in}:$, where $B^{in}$ is the bound state $in$ field.
The
coefficient of this term serves as the bound-state wavefunction; indeed as
shown
above in the
nonrelativistic case it is precisely the Schr\"odinger wavefunction.  Inserting
the two terms of the Haag expansion that have been introduced into the equation
of motion and re-normal ordering and keeping the coefficients of the
bound-state
term leads to a linear integral equation for the bound-state wavefunction.
Because the interaction is a contact interaction, this equation can be solved
exactly.  For the $J^{PC}=0^{-+}$ state, the analog of the pion, the mass is
zero, as expected from the Nambu-Goldstone theorem.  For other states, the
results agree generally with previous calculations, except in some cases the
limits on the masses differ, perhaps because the present calculation includes
crossed graphs.

{\bf 5. FINITE TEMPERATURE FIELD THEORY}

\begin{flushleft}
{\bf 5a.  Sketch of thermo field theory using the Haag expansion}
\end{flushleft}

I illustrate the application of the Haag expansion
to the solution of second quantized field
theories at finite temperature using the BCS model of
superconductivity.  In order to have a state annihilated by the annihilation
operators, which is necessary for normal ordering, I use the thermo field
theory formalism of Umezawa and
collaborators\cite{u1,u2}.  This formalism uses a doubled set of
operators to
account for finite temperature.  The annihilation
and creation operators are subjected to two Bogoliubov transformations:  one
comes from the dynamics of the electron pair interaction; the other from the
thermo field formalism which takes account of the finite temperature.  In
lowest
approximation, the method leads to the usual gap equation.
The asymptotic fields whose annihilation parts annihilate the vacuum at zero
temperature no longer annihilate the state which is a thermal mixture at finite
temperature $T$.
Indeed, no set of annihilation operators annihilates the mixed
state at finite $T$.  In order to obtain a state which is annihilated by the
annihilation parts of a suitable set of asymptotic fields, the Hilbert space of
states must be enlarged to include hole states in the thermal equilibrium state
at a given temperature and the
set of operators must include operators which annihilate and create holes in
addition to the operators which annihilate and create particles.  Thermo field
theory does this, for example for a Hilbert space with a discrete energy
eigenstate basis \{$|n \rangle$\}, by replacing the Hilbert
space of the system under consideration with the tensor product Hilbert space
with basis \{$|n \rangle \otimes |\tilde{n} \rangle$\}.  Correspondingly
the set of operators with respect to which the vacuum $| 0 \rangle \otimes
|\tilde{0} \rangle$ is cyclic is doubled and includes the particle
\{$a_k \otimes {\bf 1},
a^{\dagger}_k \otimes {\bf \tilde{1}}$\} and hole
\{${\bf 1}\otimes \tilde{a}_k, {\bf 1}\otimes\tilde{a}^{\dagger}_k$\}
operators.
(To simplify the notation, I will drop the $\otimes \bf \tilde{1}$
and $\bf{1} \otimes$ factors.)
A Bogoliubov transformation relates the original annihilation and creation
operators for the particles
together with the ``tilde'' operators which describe the holes
to another doubled set of operators whose
annihilation parts
annihilate the pure state (called the ``thermal vacuum'')
in the enlarged Hilbert space which represents the
thermal mixture in the usual theory.

Let the density operator be
\begin{equation}
\rho=Z(\beta)^{-1}e^{-\beta H}=
Z(\beta)^{-1}\sum e^{-\beta E_n}|n\rangle \langle n|,
\end{equation}
\begin{equation}
Z(\beta)=tr e^{-\beta H}=\sum e^{-\beta E_n}.
\end{equation}
Here $H$ can either be the Hamiltonian $H$ or $H- \mu N$, where $\mu$ is the
chemical potential and $N$ is the number operator.
Now consider an enlarged Hilbert space in which the tensor product basis
$\{|n \rangle \otimes |\tilde{n} \rangle\}$ replaces the basis $\{|n \rangle\}$
of the original Hilbert space.  Let the thermal vacuum be
\begin{equation}
|O(\beta)\rangle=Z(\beta)^{-1/2}\sum e^{-\beta E_n/2}
|n \rangle \otimes |\tilde{n} \rangle.
\end{equation}
Let the doubled set of operators,
$c$, $c^{\dagger}$,
$\tilde{c}$ and $\tilde{c^{\dagger}}$ be operators for the normal
modes of the total system.  The thermal vacuum at finite temperature is the
state which satisfies
\begin{equation}
c_{ki}|O(\beta)\rangle=0,~~\tilde{c}_{ki}|O(\beta)\rangle=0.
\end{equation}
The existence of a state $|O(\beta)\rangle$ which is annihilated by the
annihilation operators is essential to defining a normal-ordered operator
product.
The average of an observable $A$ in the thermal mixture at inverse
temperature $\beta$ is given by the matrix element of the corresponding
operator
$A \otimes {\bf \tilde{1}}$ in the thermal vacuum $|O(\beta)\rangle$,
\begin{equation}
tr(\rho A)=\langle O(\beta)|A \otimes {\bf \tilde{1}}|O(\beta)\rangle.
\end{equation}
For the Fermi case of interest for the electrons in superconductivity, the
Bogoliubov transformation between the operators, $b$ and $b^{\dagger}$, for
the normal modes of the electrons
in the system and the operators, $\tilde{b}$ and $\tilde{b^{\dagger}}$,
for the normal modes
of the holes on the one hand and the doubled set of operators,
$c$, $c^{\dagger}$,
$\tilde{c}$ and $\tilde{c^{\dagger}}$ which are the normal
modes of the total system on the other hand is
\begin{equation}
\left ( \begin{array}{c}
b^{\dagger}_{k1}\\
\tilde{b}_{k1}\\
\end{array} \right) =
\left ( \begin{array}{cc}
\sqrt{1-n_k} & \sqrt{n_k}\\
-\sqrt{n_k} & \sqrt{1-n_k}\\
\end{array} \right)
\left ( \begin{array}{c}
c^{\dagger}_{k1}\\
\tilde{c}_{k1}\\
\end{array} \right),
\end{equation}
\begin{equation}
\left ( \begin{array}{c}
b_{k2}\\
\tilde{b}_{k2}^{\dagger}\\
\end{array} \right) =
\left ( \begin{array}{cc}
\sqrt{1-n_k} & -\sqrt{n_k}\\
\sqrt{n_k} & \sqrt{1-n_k}\\
\end{array} \right).
\left ( \begin{array}{c}
c_{k2}\\
\tilde{c}_{k2}^{\dagger}\\
\end{array} \right)
\end{equation}
The requirement that
\begin{equation}
\langle O(\beta)|b_{k1}^{\dagger}b_{k1}|O(\beta)\rangle
=\langle O(\beta)|b_{k2}^{\dagger} b_{k2}| O (\beta)\rangle = n_{k}
=1/(e^{\beta E_k}+1)
\label{n}
\end{equation}
fixes the coefficients in the Bogoliubov transformation.  As indicated in Eq.
(\ref{n}), I assume that $n_k$
is independent of spin polarization.

The electron-electron interaction which leads to superconductivity leads to
another Bogoliubov transformation which has the form
\begin{equation}
\left( \begin{array}{c}
a^{\dagger}_{k \uparrow}\\
a_{k \downarrow}\\
\end{array} \right) =
\left( \begin{array}{cc}
u_{k} & v_{k}\\
-v^{*}_{k} & u^{*}_{k}\\
\end{array} \right)
\left( \begin{array}{c}
b^{\dagger}_{k1}\\
b_{k2}\\
\end{array} \right)
\label{b}
\end{equation}
and
\begin{equation}
\left( \begin{array}{c}
\tilde{a}_{k \uparrow}\\
\tilde{a}_{k \downarrow}^{\dagger}\\
\end{array} \right) =
\left( \begin{array}{cc}
u_{k} & v_{k}\\
-v^{*}_{k} & u^{*}_{k}\\
\end{array} \right)
\left( \begin{array}{c}
\tilde{b}_{k1}\\
\tilde{b}_{k2}^{\dagger}\\
\end{array} \right).
\end{equation}
The solution of the operator equations of motion for the $a$ and $\tilde{a}$
operators determine the coefficients $u$ and $v$ in Eq. (\ref{b});
the operators
$b$ and $\tilde{b}$ are then expressed in terms of the $c$ and $\tilde{c}$ set.
The final result gives the $a$ and $\tilde{a}$ operators in terms of the
$c$ and $\tilde{c}$ operators,
\begin{equation}
\left ( \begin{array}{c}
a^{\dagger}_{k \uparrow}\\
a_{-k \downarrow}\\
\tilde{a}_{k \uparrow}\\
\tilde{a}^{\dagger}_{-k \downarrow}\\
\end{array} \right) =
\left ( \begin{array}{cccc}
u_{k}\sqrt{1-n_{k}} & v_{k}\sqrt{1-n_{k}} & u_{k}\sqrt{n_{k}} & -v_{k}\sqrt
{n_{k}}\\
-v^{*}_{k}\sqrt{1-n_{k}} & u^{*}_{k}\sqrt{1-n_{k}} & -v^{*}_{k}\sqrt{n_{k}} &
-u^{*}_{k}\sqrt{n_{k}}\\
-u_{k}\sqrt{n_{k}} & v_{k}\sqrt{n_{k}} & u_{k}\sqrt{1-n_{k}} & v_{k}\sqrt{1-n
_{k}}\\
v^{*}_{k}\sqrt{n_{k}} & u^{*}_{k}\sqrt{n_{k}} & -v^{*}_{k}\sqrt{1-n_{k}} &
u^{*}_{k}\sqrt{1-n_{k}}\\
\end{array} \right)
\left ( \begin{array}{c}
c^{\dagger}_{k1}\\
c_{k2}\\
\tilde{c}_{k1}\\
\tilde{c}^{\dagger}_{k2}\\
\end{array} \right).
\label{in}
\end{equation}

{\bf 5b. Solution of the finite temperature equations of motion}

Assume there is an attractive interaction between
pairs of electrons with opposite spin and momenta which acts only on electrons
near the Fermi surface\cite{bcs}.
The total Hamiltonian is
\begin{equation}
\bar{H}=H-\tilde{H},
\end{equation}
\begin{equation}
H=\Sigma_{ks}(k^2/2m)a^{\dagger}_{ks}a_{ks}+(1/4)\Sigma_{k\ell}
[a^{\dagger}_{k\uparrow}, a^{\dagger}_{-k\downarrow}]V_{k\ell}
[a_{-\ell\downarrow}, a_{\ell\uparrow}],
\end{equation}
where the potential acts only near the Fermi level,
\begin{equation}
V_{kl} =
\left \{ \begin{array}{ccc}
-V_0, & |\hbar^2 p^2/2m-\mu|\leq \Delta \epsilon, & p = k {\rm ~~or~~} l\\
0, &~~  {\rm otherwise}. & \\
\end{array}
\right.
\end{equation}
The tilde Hamiltonian, $\tilde{H}$, is the same as $H$, except that $\tilde{a}$
and $\tilde{a^{\dagger}}$ replace $a$ and $a^{\dagger}$.

It is convenient to measure energy relative to the Fermi surface and to use
$H- \mu N$ instead of $H$ as the energy.  This has the effect of changing the
coefficient of the kinetic term in the Hamiltonian from ${\bf k}^2/2m$ to
$\hat{\epsilon_k}={\bf k}^2/2m-\mu$.  I retain the symbol $H$ for this new
``Hamiltonian'' and the symbol $\bar{H}$ for the new ``total Hamiltonian.''

The time dependence of any operator is
${\cal{O}}(t)=e^{i \bar{H} t}{\cal{O}}e^{-i \bar{H} t}$.
Thus the equation of motion
is
\begin{equation}
-i \partial_t {\cal{O}}(t)=[\bar{H}, {\cal{O}}(t)].
\end{equation}
The equations of motion for creation and annihilation operators are
\begin{equation}
-i\partial_{t}a^{\dagger}_{k\uparrow}
=\hat{\epsilon_{k}}a^{\dagger}_{k\uparrow} +
(1/2)\Sigma_{\ell}[a^{\dagger}_{\ell\uparrow}, a^{\dagger}_{-\ell
\downarrow}]V_{\ell k}a_{-k\downarrow},
\end{equation}
\begin{equation}
-i\partial_{t}a_{-k\downarrow}
=-\hat{\epsilon_{k}}a_{-k\downarrow}+(1/2)\Sigma_\ell
a^{\dagger}_{k\uparrow}V_{k\ell}[a_{-\ell\downarrow},a_{\ell\uparrow}],
\end{equation}
\begin{equation}
-i\partial_{t}\tilde{a}_{k\uparrow}
=\hat{\epsilon_{k}}\tilde{a}_{k\uparrow}+(1/2)
\Sigma_{\ell}\tilde{a}^{\dagger}_{-k\downarrow}V_{k\ell}[\tilde{a}_{-\ell
\downarrow}, \tilde{a}_{\ell\uparrow}],
\end{equation}
\begin{equation}
-i\partial_{t}\tilde{a}^{\dagger}_{-k\downarrow}
= -\hat{\epsilon_{k}}\tilde{a}
^{\dagger}_{-k\downarrow}+(1/2)\Sigma_{\ell}[\tilde{a}^{\dagger}
_{\ell\uparrow},\tilde{a}^{\dagger}_{-\ell\downarrow}] V_{\ell k}\tilde{a}
_{k\uparrow}.
\end{equation}
Use a Haag expansion in the thermal $in$ (or $out$) fields
to get an approximate solution of the operator equations of motion.  In the
lowest approximation, use Eq.(\ref{in}), insert it in the equations of motion,
re-normal order and keep only the linear terms in the $c$ operators
to get equations for the unknown coefficients $u_k$ and
$v_k$.   The result has the form
\begin{equation}
\sum_j(F_k+\sum_lG_{kl})_{ij}{\cal C}_j=0,
\end{equation}
where ${\cal C}_{j}=(c_{k1}^{\dagger}, c_{k2}, {\tilde c}_{k1},
{\tilde c}_{k2}^{\dagger})$ (as a column vector).  Since the $c$ and
$\tilde{c}$
operators are linearly dependent, each of the 16 equations,
\begin{equation}
(F_k+\sum_l G_{kl})_{ij}=0, i,j=1 ~{\rm to }~ 4,
\end{equation}
must hold.  Only two of these equations,
\begin{equation}
(E_{k} - \hat{\epsilon_{k}} - V_{kk} w_{k}) u_{k} +
\Sigma_{\ell}V_{k \ell}y_{\ell}v_{k} = 0,
\end{equation}
\begin{equation}
(E_{k} + \hat{\epsilon_{k}} + V_{kk} w_{k}) v_{k}
+ \Sigma_{\ell}V_{k \ell}y_{\ell}u_{k} = 0,
\end{equation}
where
\begin{equation}
w_k \equiv v^2_k(1-n_k)+u_k^2n_k, ~~~y_l \equiv u_lv_l(1-2n_l),
\label{wy}
\end{equation}
are linearly independent.  It is convenient to write these equations in
matrix form and to make the equations appear simpler by introducing new
symbols
\begin{equation}
\epsilon_k \equiv \hat{\epsilon_k}+V_{kk}w_k,
{}~~~\Delta_k \equiv \sum_l V_{kl}y_l.
\label{ab}
\end{equation}
Then
\begin{equation}
\left  ( \begin{array}{cc}
E_k-\epsilon_k   &   \Delta_k  \\
\Delta_k       &   E_k+\epsilon_k\\
\end{array} \right )
\left ( \begin{array}{c}
u_k\\
v_k\\
\end{array}  \right )=0.
\label{matrix}
\end{equation}
There is a solution only if the determinant of the matrix vanishes, i.e., if
\begin{equation}
E_k^2=\epsilon_k^2+\Delta_k^2.
\label{det}
\end{equation}
The solution, which is
implicit because $\epsilon$ and $\Delta$
depend on $u_k$ and $v_k$ via Eq.(\ref{wy}) and
(\ref{ab}), for $u_k$ and $v_k$ is
\begin{equation}
u_k=\sqrt{(E_k+\epsilon_k)/2E_k}, ~~ v_k=-\Delta_k/\sqrt{2E_k(E_k+\epsilon_k)}.
\label{uv}
\end{equation}
It is easy to check that this solution satisfies the constraint
$u_k^2+v_k^2=1$.
When Eq.(\ref{uv}) is inserted in the definition of $\Delta$, the celebrated
gap
equation results,
\begin{equation}
\Delta_k=\sum_l V_{kl} (\Delta_l/2 E_l)(1-2n_l).
\label{gap1}
\end{equation}
With the choices,
\begin{equation}
V_{kl} =
\left \{ \begin{array}{ccc}
-V_0, & |\epsilon_k|\leq \Delta \epsilon, & p = k {\rm ~~or~~} l\\
0, &~~  {\rm otherwise}, & \\
\end{array}
\right.                             \label{pot}
\end{equation}
\begin{equation}
\Delta_k =
\left \{ \begin{array}{ccc}
\Delta(T), & |\epsilon_k|\leq \Delta \epsilon, \\
0, &~~  {\rm otherwise}, & \\
\end{array}
\right.                           \label{Delta}
\end{equation}
using Eq.(\ref{n}) and making the usual
replacement of the sum by an integral,
the gap equation becomes
\begin{equation}
V_0N(0) \int_{0}^{\Delta \epsilon} d \epsilon_l~~ tanh (\beta
E_l/2)/E_l=1,
\label{gap2}
\end{equation}
where $E_k=\sqrt{\epsilon_k^2+\Delta^2(T)}$.
This is the usual gap equation for this model and the usual results for
$\Delta(0)/kT_c=1.764$, $N(0)V_0~ln(1.13\Delta \epsilon/kT_c)=1$, etc.,
where $T_c$ is the transition temperature, follow.

{\bf 6. THE SCHWINGER MODEL}

The Schwinger model\cite{sw,ab} is
massless two-dimensional quantum electrodynamics, an exactly soluble model.
In the
Lorentz gauge,\footnote{This form of the Lagrangian ignores wavefunction
renormalization of the spinor field.  Taking account of this wavefunction
renormalization, the spinor term in the Lagrangian is
$\bar \psi_u i D\!\!\!\! / \psi_u
-\langle \bar \psi_u i D\!\!\!\! / \psi_u \rangle_0=Z\bar \psi i D\!\!\!\! /
\psi$, where $\psi_u$ is the unrenormalized spinor field and
$Z$ is the spinor wavefunction renormalization.  We suppress the
factor $Z$ below.}
\beee
{\cal L}=\bar \psi i D\!\!\!\! / \psi
-\frac{1}{4} F_{\mu \nu} F^{\mu \nu}-\frac{1}{2}(\partial \cdot A)^2,\label{1}
\eeee
where
\beee
D^{\mu}=\partial^{\mu} -e A^{\mu},~~
F^{\mu \nu}=\partial^{\mu}A^{\nu}-\partial^{\nu}A^{\mu}. \label{2}
\eeee

Lowenstein and
Swieca\cite{ls} found
an operator ansatz that yields the matrix elements computed by Schwinger.
Their
ansatz solution has the following form:
\beee
A^{\mu}(x)=-\sqrt{\frac {\pi}{e^2}} (\epsilon^{\mu \nu}\partial_{\nu} \Sigma
+\partial^{\mu} \eta), \label{3}
\eeee
\beee
\psi(x)=:\exp [i \sqrt{\pi} ( \gamma_5 \Sigma(x) - \eta(x))]:\psi^{(0)}(x),
\label{4}
\eeee
where $\eta$ is a free neutral massless field with negative metric
corresponding to the gauge degrees
of freedom satisfying
$[\eta(x),\dot \eta(y)]_{ET}=-i\delta(x^1-y^1)$ , while $\Sigma$ is a free
neutral
massive field with positive metric
representing the physical degrees of freedom satisfying
$[\Sigma(x),\dot \Sigma(y)]_{ET}=i\delta(x^1-y^1)$ and
$\psi^{(0)}$ is a solution of the free massless
Dirac equation.  (ET stands for equal
time.)
This solution displays the main property of the Schwinger model: the only
physical state is a free particle of mass $e /\sqrt{\pi}$, but the solution
does
not obey the canonical commutation relations for $A_{\mu}$.  The Haag expansion
provides a solution that does obey these relations\cite{gat}.

Use the Lagrangian Eq.(\ref{1}), but drop surface terms so that the gauge
field part becomes
\beee
-\frac{1}{2}(\partial_{\mu}A_{\nu})(\partial^{\mu}A^{\nu}). \label{6}
\eeee
Since the Lorentz group (without inversions) is abelian in 1+1, all irreducible
representations are one-dimensional; thus the vector and spinor fields as to
in the
model are composed of one-dimensional irreducibles arbitrarily pasted together.
Express the Lagrangian in terms of the irreducible fields in the
basis in which
\beee
A^0=\frac{1}{2}(A^++A^-),~~A^1=\frac{1}{2}(A^+-A^-), ~~\psi=(\psi_1,\psi_2),
\label{7}
\eeee
with
\beee
\gamma^0=\left( \begin{array}{cc}
               0&1\\
               1&0
                \end{array} \right),~~
\gamma^1=      \left( \begin{array}{cc}
               0&-1\\
               1&0
                \end{array} \right),~~
\gamma^5=      \left( \begin{array}{cc}
               1&0\\
               0&-1
                \end{array} \right).~~
\label{8}
\eeee
In terms of the irreducible fields,
\beee
{\cal L}=\psi_1\dggg(2i\partial^+ -eA^-)\psi_1+
\psi_2\dggg(2i\partial^- -eA^+)\psi_2+\frac{1}{2}(\partial^1A^+
\partial^1A^- - \partial^0A^+\partial^0A^-). \label{9}
\eeee
using lightcone coordinates, $x^+=x^0+x^1,~~x^-=x^0-x^1$.
The corresponding derivatives are
$\frac{\partial}{\partial x^{\pm}}=\frac{1}{2}(\frac{\partial}
{\partial x^0} \pm \frac{\partial}{\partial
x^1})$.  Define these so that $\frac{\partial}{\partial x^{\pm}} x^{\pm}=1$.
Note that although the fields $A^{\pm}$ are lightcone fields, I am {\it not}
using lightcone quantization, but rather am using equal-time canonical
quantization.  The naive operator equations of motion are
\beee
\Box A^+- 2e\psi\dggg_1\psi_1=0, \label{10}
\eeee
\beee
\Box A^-- 2e\psi\dggg_2\psi_2=0, \label{11}
\eeee
\beee
(2i\partial^+-eA^-)\psi_1=0, \label{12}
\eeee
\beee
(2i\partial^--eA^+)\psi_2=0. \label{13}
\eeee
As Schwinger pointed out in his original paper, the spinor bilinear products
require a line integral of the ``vector'' potential
between the $\psi\dggg$ and the $\psi$ in order to ensure gauge invariance;
this
is done explicitly below using point-splitting.
For example, $\psi_1\dggg \psi_1$ is replaced
by
\beee
\lim_{\epsilon \rightarrow 0} \, \frac{1}{2}
\left[
\psi\dggg_1(x+\epsilon)e^{-ie \int^{x+\epsilon}_{x} A_{\mu}(w) dw^{\mu}}
 \psi_1(x)+ cc. \right].  \label{14}
\eeee
The canonical momenta are
\beee
\pi_{A^+}=-\frac{1}{2}\partial^0A^-,~~\pi_{A^-}=-\frac{1}{2}\partial^0A^+,~~
\pi_{\psi_j}=i \psi_j\dggg.
\label{15}
\eeee

Solve the Dirac equations by exponentiation,
\beee
\psi_1(x)={\cal P}exp[-\frac{ie}{2}\int^{x^+}_{-\infty}A^-(w^+,x^-)dw^+]
\psi_1^{(0)}(x^-),~~\partial^+\psi_1^{(0)}=0, \label{16}
\eeee
\beee
\psi_2(x)={\cal P}exp[-\frac{ie}{2}\int^{x^-}_{-\infty}A^+(x^+,w^-)dw^-]
\psi_2^{(0)}(x^+),~~\partial^-\psi_2^{(0)}=0. \label{17}
\eeee
The point-splitting vector is taken spacelike, $\epsilon=(0,\epsilon^1),
{}~~\epsilon^{\pm}=\pm \epsilon^1$.  Thus, for example, $\psi_1\dggg$ must be
replaced by
\beee
\psi\dggg_1(x+\epsilon)=\psi^{0~\dagger}_1(x^--\epsilon^1)\bar{\cal P}
exp[\frac{ie}{2}\int^{x^++\epsilon^1}_{-\infty}
A^-(w^+,x^--\epsilon^1)dw^+].  \label{18}
\eeee
The symbols ${\cal P}$ and $\bar{\cal P}$ stand for path and antipath ordering,
respectively.
The result of the point-splitting differs from the usual one by having
integrated (nonlocal) terms.
The equations for $A^{\pm}$ become
\beee
(\Box + \frac{e^2}{2 \pi})A^+-\frac{e^2}{2 \pi}\int^{x^+}_{-\infty}
\frac{\partial A^-}{\partial x^-}(w^+,x^-)dw^+=2e\psi^{(0)\dagger}_1(x^-)
\psi^{(0)}_1(x^-) \label{19}
\eeee
\beee
(\Box + \frac{e^2}{2 \pi})A^--\frac{e^2}{2 \pi}\int^{x^-}_{-\infty}
\frac{\partial A^+}{\partial x^+}(x^+,w^-)dw^-=2e\psi^{(0)\dagger}_2(x^+)
\psi^{(0)}_2(x^+). \label{20}
\eeee
The integrated terms here can be removed by taking derivatives with respect
to the upper limit.  Combining the resulting equations leads to
\beee
\Box~~ \partial \cdot A=0 \label{21}
\eeee
\beee
(\Box+\frac{e^2}{\pi})~\epsilon_{\mu \nu}\partial^{\mu}A^{\nu}=0; \label{22}
\eeee
thus $\partial \cdot A \equiv \eta$ is a massless field and
$\epsilon_{\mu \nu}\partial^{\mu}A^{\nu} \equiv \Sigma$ is a field of
mass $e/\sqrt{\pi}$.  The fields $\eta$ and $\Sigma$
(which is the electric field in 1+1 dimensions) are the
gauge-variant and gauge-invariant
degrees of freedom, respectively.
Then $\Box~ A^{\mu}$ must be a linear combination of
$\partial^{\mu} \eta$ and $\epsilon^{\mu \nu} \partial_{\nu} \Sigma$.
$A^{\mu}$ must be the convolution of the
$\bar{\Delta}(x)=-\frac{1}{2}\epsilon(x^0)\Delta(x)$ Green's function with
this linear combination plus terms annihilated by $\Box$.
The convolution of
$\bar{\Delta}(x)$
with $\eta$ does not exist, because, formally, it is $\int d^2y \bar
\Delta(x-y)\eta(y) =\int d^2k exp(-ik \cdot x)\delta(k^2) \tilde \eta(k)/k^2$,
which is ill-defined.  Because of this, a new
field $a$ that obeys $\Box~ a =\eta$ must be introduced.  This was  first
done by Capri and Ferrari\cite{cf}.
Thus
\beee
A^{\mu}=c_1\partial^{\mu}a+c_2 \epsilon^{\mu \nu}\partial_{\mu}\Sigma
+c_3 \partial ^{\mu} \eta +c_4 \bar{\psi}^{(0)}\gamma^{\mu}\psi^{(0)}.
\label{23}
\eeee
For the massless case,
\beee
\bar{\psi}^{(0)}\gamma^{\mu}\psi^{(0)}=\partial^{\mu} \phi,~~\Box ~\phi=0,
\label{24}
\eeee
where $\phi$ is a free positive-metric scalar field.  Using
\beee
[\eta,\dot{\eta}]_{ET}=\tau i \delta,~~[\eta,\dot{a}]_{ET}=c_{a\eta} i
\delta,~~
[a,\dot{\eta}]_{ET}=c_{a\eta} i \delta,~~[a,\dot{a}]_{ET}=c_{aa} i \delta.~~
\label{33}
\eeee
with the choices $c_{a \eta}=c_{aa}=0$ and
\beee
\tau=-1,~~c_1=\mp \sqrt{\frac{e^2}{5 \pi}},~~c_2=\pm \sqrt{\frac{\pi}{e^2}},~~
c_3=\pm \sqrt{\frac{5 \pi}{e^2}}. \label{34}
\eeee
yields a solution in which the vector potential obeys the canonical commutation
relations.

The canonical commutation relations in field theory and their predecessors in
classical mechanics and quantum mechanics are important for many reasons.
The Poisson brackets in classical mechanics, for example,
ensure
that the Hamiltonian is the generator of time translations.  In
quantum mechanics, for example,
the relation $[x,p]=i\hbar$ leads to the uncertainty relation.
In quantum field
theory, the CCR's lead to the free field being a collection of quantized
oscillators.  In nonrelativistic field theories at least, the CCR's imply
unitarity\cite{cg}.  A new feature of the canonical commutation
relations in quantum field
theory is that they ensure that the asymptotic fields have the proper free
commutation relation.  (The renormalized canonical commutation relations will
do
as well as the original CCR's for this purpose.)
For these reasons, a solution that obeys the canonical commutation
relations is important.

{\bf 7. SUMMARY AND OUTLOOK FOR FUTURE WORK}

The N quantum approach using the Haag expansion has met the test of non-gauge
theories, including bound states and both spontaneous and dynamical symmetry
breaking.  Previous work by Amit Raychaudhuri\cite{r} and work presently in
progress in collaboration with Eli
Hawkins\cite{hg} and with
Rashmi Ray and Felix
Schlumpf\cite{grs} in both non-gauge theories and in
non-confining gauge theories give promise of
using the N quantum approach to treat bound states in a way that has advantages
over the Bethe-Salpeter equation.  The treatment of confined degrees of freedom
in quantum chromodynamics remains a goal for the future.

{\bf ACKNOWLEDGEMENTS}

It is a pleasure to thank Masud Chaichian for inviting me to
participate in this workshop and for his hospitality in Helsinki and in
Saariselka.  I also thank my recent and present collaborators, Gil Gat, Eli
Hawkins, Rashmi Ray and Felix Schlumpf.  I am indebted to Joe Sucher for a
careful reading of the manuscript that resulted in significant improvements in
the exposition.

\end{document}